\documentclass[manuscript]{aastex}

\shorttitle{UV Color Variability of Luminous QSOs}
\shortauthors{Sakata et al.}

\begin{document}

\title{UV Continuum Color Variability of Luminous SDSS QSOs}
\author{Yu Sakata\altaffilmark{1,2}, Tomoki Morokuma\altaffilmark{1,4,5}, Takeo Minezaki\altaffilmark{1}, Yuzuru Yoshii\altaffilmark{1,3},\\ Yukiyasu Kobayashi\altaffilmark{4}, Shintaro Koshida\altaffilmark{2}, and Hiroaki Sameshima\altaffilmark{1,2}} 

\altaffiltext{1}{Institute of Astronomy, School of Science, University of Tokyo, 2-21-1 Osawa, Mitaka, Tokyo 181-0015, Japan; yusakata@ioa.s.u-tokyo.ac.jp.}
\altaffiltext{2}{Department of Astronomy, School of Science, University of Tokyo, 7-3-1 Hongo, Bunkyo-ku, Tokyo 113-0013, Japan.}
\altaffiltext{3}{Research Center for the Early Universe, School of Science, University of Tokyo, 7-3-1 Hongo, Bunkyo-ku, Tokyo 113-0033, Japan.}
\altaffiltext{4}{National Astronomical Observatory, 2-21-1 Osawa, Mitaka, Tokyo 181-8588, Japan.}

\altaffiltext{5}{Research Fellow of the Japan Society for the Promotion of Science}
\begin{abstract}

We examine whether the spectral energy distribution of UV continuum emission of 
active galactic nuclei (AGNs) changes during flux variation. 
We used multi-epoch photometric data of QSOs in the Stripe 82
observed by the Sloan Digital Sky Survey (SDSS) Legacy Survey 
and selected 10 bright QSOs observed with high photometric accuracies,
in the redshift range of $z=1.0-2.4$ 
where strong broad emission lines such as Ly$\alpha $ and \ion{C}{4}
do not contaminate SDSS filters, to examine 
spectral variation of the UV continuum emission
with broad-band photometries.
All target QSOs showed clear flux variations
during the monitoring period 1998$-$2007,
and the multi-epoch flux data in two different bands
obtained on the same night showed a linear flux-to-flux relationship
for all target QSOs.
Assigning the flux in the longer wavelength to the x-axis 
in the flux-to-flux diagram, the x-intercept of the best-fit 
linear regression line was positive for most targets, 
which means that their colors in the observing bands 
become bluer as they become brighter. 
Then, the host galaxy flux was estimated
on the basis of the correlation between the stellar mass of the bulge
of the host galaxy and the central black hole mass; 
the latter was estimated on the basis of the luminosity scaling relations
for \ion{C}{4} or \ion{Mg}{2} emission lines and their line width.
We found that the longer wavelength flux of the host galaxy was 
systematically smaller than that of the fainter extension of 
the best-fit regression line at the same shorter wavelength flux for most targets. 
This result strongly indicates that
the spectral shape of the continuum emission of QSOs
in the UV region ($\sim $1400$-$3600\AA \, in rest-frame wavelength)
usually becomes bluer as it becomes brighter.
The multi-epoch flux data in the flux-to-flux diagram
were found to be consistent with the wavelength-dependent
amplitude of variation presented in \citet{2004ApJ...601..692V},
which showed a larger amplitude of variation in shorter wavelengths.
We also found that
the multi-epoch flux-to-flux plots
could be fitted well with the standard accretion disk model
changing the mass accretion rate with a constant black hole mass
for most targets. 
This finding strongly supports the standard accretion disk model
for UV continuum emission of QSOs.

\end{abstract}

\section{Introduction}
\label{s_intro}

It is generally accepted that the vast amount of continuum emission of
type 1 active galactic nuclei (AGNs) in UV and optical wavelengths
originates in the accretion disk surrounding a supermassive black hole,
and the UV-optical variability found at the beginning of AGN studies
has been considered a powerful tool for understanding
the nature of the AGN central engine.  
However, the mechanism of this variability is still under discussion.
Many models for variability have been proposed, including
accretion disk instabilities \citep{1984ARA&A..22..471R, 1998ApJ...504..671K},
X-ray reprocessing (Krolik et al. 1991; Kawaguchi in prep.),
star collisions (Courvoisier, Paltani, \& Walter 1996; Torricelli-Ciamponi et al. 2000),
and gravitational microlensing (Hawkins 1993); 
however, none of these successfully explained more than a few properties
of UV-optical variability \citep{2004ApJ...601..692V}. 

The spectral variability of the UV-optical continuum emission
during flux variation
is a key property for understanding the central engines
and their variability mechanisms in AGNs.
For example,
since the spectral energy distribution (SED)
of a hot spot or a flare in the accretion disk
caused by local enhancement of mass accretion
or disk instabilities would be different from
that of the entire disk,
the spectral shape of continuum emission
is expected to vary with the flux variation
when caused by those mechanisms.
On the other hand,
a change of the global mass accretion rate of the disk
and a certain reprocessing model \citep{1999MNRAS.302L..24C}
would not change the temperature distribution
of the accretion disk at larger radii; 
thus, the absence of the spectral variation
of optical continuum emission during flux variation
suggests those variation mechanisms.

Sakata et al. (2010; hereafter Paper I)
addressed the spectral variability of optical continuum emission of AGNs.
Paper I examined the long-term multi-band monitoring data of 11 nearby AGNs 
and precisely estimated the contaminated flux of the host galaxy
and the narrow emission lines.
Then, it was found that 
the multi-epoch optical flux data in any two different bands
obtained on the same night showed a very tight linear flux-to-flux relationship
for all target AGNs 
and that the non-variable component of the host galaxy plus narrow lines
was located on the fainter extension of the linear regression line
of multi-epoch flux-to-flux plots.
From these results, Paper I concluded that the spectral shape
of AGN continuum emission in the optical region ($\sim4400-7900$\AA)
does not systematically change during flux variation 
and that the trend of spectral hardening
in which the optical continuum emission becomes bluer as it becomes brighter,
which has been reported by many studies
\citep{1990ApJ...354..446W, 1999MNRAS.306..637G, 2000ApJ...540..652W}, 
is caused by the contamination of the non-variable component
of the host galaxy plus narrow emission lines, 
which is usually redder than AGN continuum emission.  

In contrast to optical continuum emission, two opposite claims have not been resolved 
for the spectral variability of UV continuum emission. 
\citet{2004ApJ...601..692V} statistically examined
the properties of flux variations of QSOs 
from the two-epoch photometric observations of
about 25,000 QSOs obtained by the Sloan Digital Sky Survey (SDSS) 
and found a larger amplitude of variation
at shorter wavelengths in the UV region of $\lambda < 4000$ \AA,
indicating spectral hardening during flux variation of QSOs.
\citet{2005ApJ...633..638W} obtained a composite
of the differential spectrum of two-epoch spectroscopic observations
for hundreds of SDSS QSOs 
and found that the composite differential spectrum
was bluer than the composite spectrum of QSOs,
which also indicates spectral hardening of the UV continuum emission.

On the other hand,
\citet{1996A&A...312...55P}
applied principal component analysis (PCA) to
the multi-epoch UV spectra of 15 nearby AGNs
obtained by the {\it International Ultraviolet Explorer} (IUE) satellite 
and concluded that the UV flux variation of the AGNs
consists of a variable component with a constant spectral shape
and a non-variable component. 
Based on the decomposition of the multi-epoch spectra to a variable 
and a non-variable components, they further concluded that 
the variable component has a power-law shape of fixed spectral index 
and that the non-variable component can be reproduced using 
the sum of a steep Balmer continuum and a \ion{Fe}{2} pseudo-continuum, 
which corresponds to the spectral feature of the small blue bump (SBB).
\citet{1995MNRAS.274....1S} and \citet{1997ApJS..110....9R}
observed NGC 4593 and Fairall 9, respectively, using the IUE satellite.
They found that the multi-epoch UV flux data in two different bands
obtained on the same night showed a linear flux-to-flux relationship 
and that the contaminated flux of the SBB was located on a fainter extension of
the linear regression line.
These authors concluded that
the UV continuum emission retains a constant spectral shape
during the AGNs' flux variation. 

In this paper,
we examine the spectral variability of the UV continuum emission of AGNs, 
in the same way as in Paper I,
on the basis of the long-term multi-epoch photometric data
of 10 mid-redshift luminous QSOs obtained by the SDSS.
In Section \ref{s_obsdata},
we describe the selection of target QSOs, their basic properties 
such as the central black hole mass and accretion rate,
and present their light curves. 
In Section \ref{s_result},
we examine the UV color variability of the target QSOs 
from the analysis of the flux-to-flux plots
and the contaminated fluxes of the host galaxies. 
In Section \ref{s_discuss},
we compare our results with previous observational studies
about the UV color variability of AGNs
and also with an accretion disk model.
In Section \ref{s_summary}, we summarize our results.
We assume cosmological parameters of 
$(h_0, \Omega _0, \lambda_0)=(0.73,0.27,0.73)$ throughout this paper.

\section{Multicolor Light Curve of QSOs}
\label{s_obsdata}

\subsection{SDSS Stripe 82 Data}
\label{s_stripe82data}

Stripe 82 is located in the South Galactic Gap
and was scanned multiple times by the SDSS Legacy Survey 
in order to enable a deep co-addition of the data and to find variable objects.
It is defined as the region spanning 8 h in right ascension (RA)
from $\alpha$ = 20$^h$ to 4$^h$ and $2^\circ.5$ in declination (Dec.)
from $\delta$ = $-1^\circ.25$ to $1^\circ.25$,
consisting of two scan regions referred to
as the north and south strips. 
Both the north and south strips have been repeatedly imaged
in $u$, $g$, $r$, $i$, and $z$ bands about 80 times on average
by more than 300 nights of observations from 1998 to 2007,
with about 70 percent imaging runs obtained after 2005
since when the SDSS-II Supernova Survey started.
There are about 37,000 QSO candidates in the Stripe 82 region
\citep{2009ApJS..180...67R} of which
spectroscopic data were available for about 8,300 candidates.

\subsection{Target Selection}
\label{s_targetselec}

In order to examine rest-UV continuum color variation of QSOs,
we first selected the spectroscopically identified QSOs in Stripe 82 
from the QSO candidate list of \citet{2009ApJS..180...67R}.
Then, we selected the targets using the following criteria: 
(1) The target redshift is $z=1.05,1.54,1.71,2.35\pm0.05$ 
in which strong broad lines such as Ly$\alpha$ and \ion{C}{4} do not contaminate 
SDSS filters and in which rest wavelengths of the $u$- or $g$-band
are just longer or shorter than 
that of the \ion{C}{4} line (around $1400$ \AA \, or $1730$ \AA).
\footnote{When the \ion{C}{4} line is included in the SDSS $u$ or $g$ bands
at a certain redshift, its flux is estimated 
as about $30-50$\% of UV continuum flux based on the typical \ion{C}{4} equivalent width
for SDSS QSOs \citep{2008MNRAS.389.1703X}.} 
(2) The signal to noise ratio (S/N) of the photometric data
in two bands used for the analysis is more than 30.
The S/N values are taken from \citet{2009ApJS..180...67R} 
and are based on the single-epoch of observation.
Finally, we selected 10 luminous QSOs.  
Although wide wavelength coverage is preferable for examining the spectral variability, 
we basically used the $i$-band data in place of the $z$-band data as the longest wavelength observation 
because the S/N value of the former is much higher than that of the latter for most QSOs.
The targets, their redshifts, and the filter selections are listed in Table \ref{t_filterwave}.

We then calculated the black hole mass and the mean Eddington ratio of
accretion rate for our targets. 
The black hole mass was estimated using the scaling relations
for the \ion{C}{4} emission line, or the \ion{Mg}{2} emission line
if the \ion{C}{4} emission line was unavailable,
according to \citet{2009ApJ...700...49T}.
The full width at half maximum (FWHM) of the emission line was estimated from
the standard deviation of the emission-line width
taken from the SDSS database,
which was derived by fitting the Gaussian profile to the emission line,
by multiplying a factor of $2\sqrt{2\ln2}$.
We selected the specific luminosity at $1350$ \AA \, 
for the scaling relation of the \ion{C}{4} emission line 
and at $3000$ \AA \,  for that of the \ion{Mg}{2} emission line
where the effective wavelengths in the rest frame of the photometric
bands are neighbored,
estimated by interpolating or extrapolating the fluxes
of the two photometric bands,
and they were averaged over the light curves of the targets.
Strong absorption features in the \ion{C}{4} emission line
were found in three targets
(J0136$-$0046, J0346$-$0042, and J2134$+$0048),
and the error of the black hole mass estimate
might be larger for them. 

The bolometric luminosity was calculated using the equation 
$L_{bol}=5\lambda L_{\lambda}(3000$ \AA$)$ \citep{2005ApJ...629...61K} 
and the Eddington luminosity was calculated as
$L_{Edd}=\frac{4\pi Gcm_p}{\sigma_e}M_{BH}$,
where $G$ is the gravitational constant, $c$ is light speed, $m_p$ is the mass of a proton, 
$\sigma_e$ is the Thomson scattering cross-section, 
and $M_{BH}$ is the black hole mass estimated as described above.
Then, the mean Eddington ratio of accretion rate, $L_{bol}/L_{Edd}$ was calculated.
The specific luminosities, the standard deviation of the emission
lines, the black hole masses, and the Eddington ratios are listed
in Table \ref{t_objbasicparm}. 
The black hole mass of the targets are about $10^{9-10}M_{\odot}$,
and the mean Eddington ratio is between $0.1-1.2$. 

We also examined the radio flux of the targets.
As listed in Table \ref{t_objbasicparm},
8 of the 10 targets were not detected by 
the Very Large Array (VLA) FIRST survey (Becker, White, \& Helfand 1995),
and the remaining two targets are not covered by
the data of the FIRST survey.
From these estimations of the black hole mass and the Eddington ratio,
and the examination of the radio flux,
the targets can be characterized as
radio quiet QSOs with very massive black holes.

\subsection{Photometry}
\label{s_gooddataselec}

We used the PSF magnitude corrected with Galactic extinction
obtained from the SDSS database for the photometry of the target QSOs 
because the photometric error of the PSF magnitude
was smaller than that of the aperture magnitude.
Then, we further examined those data in order to improve
the photometric accuracy of the light curves of the targets.

First, we evaluated the atmospheric condition for photometry
using the QA value and selected the data after 2004,
when the QA value was available. It was provided by the SDSS database
and gives $1\sigma$ fluctuations by the millimag in the recalibrated
g-band camcol 3 zero point for fields in the run.
If this number is zero, then the night was photometric or
no recalibration was done,
and if this number is non-zero, 
a large number corresponds to more variable clouds
and worse photometric calibration can be expected
in general.
Then, we adopted only the data whose QA values were less than 50,
that is, $0.05$ mag error of the photometric calibration,
if the QA value was available.

In addition, we examined the atmospheric condition
using the flux of reference stars located near the target,
which should be constant.
When the flux of the reference star for a target QSO at some epoch
was more or less than $3\sigma$ from its mean flux for all available data,
we did not use the photometric data of the target at that epoch.
Even though the QA value was unavailable for the data before 2004,
the selection procedure based on the flux of the reference stars
efficiently rejected the data obtained in  bad atmospheric condition.

Finally, the magnitude of the target QSO was measured
relative to the nearby reference stars,
then the magnitudes of the reference stars are added.
One or two reference stars were selected, 
which locate in the same field of the target
and are brighter than the target by more than 2 mag.
This relative photometry would reduce the fluctuation
of the flux caused by changing atmospheric conditions
during the observation and would improve the photometric accuracy
of the light curves of the targets.

\begin{figure}
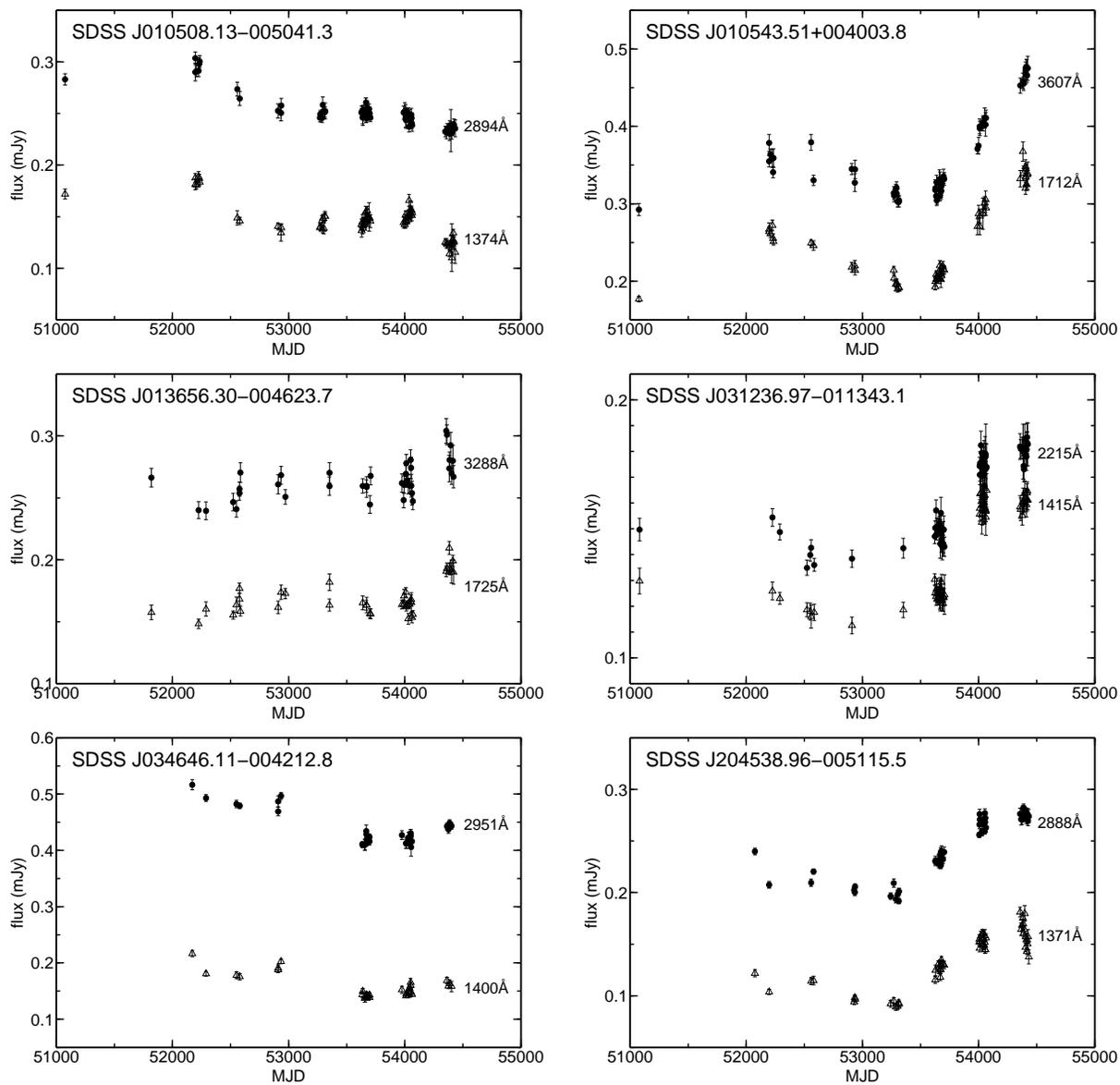

\plottwo{fig1.eps}{fig2.eps}

\plottwo{fig3.eps}{fig4.eps}

\plottwo{fig5.eps}{fig6.eps}

\caption{Light curves in the two SDSS filters (Table \ref{t_filterwave}) 
for SDSS J010508.13-005041.3, SDSS J010543.51+004003.8, SDSS J013656.30-004623.7, 
SDSS J031236.97-011343.1, SDSS J034646.11-004212.8, and SDSS J204538.96-005115.5. 
The wavelengths provided in the figures are the rest-frame effective wavelengths
of the two filters.
\label{f_lc1}
}
\end{figure}

The light curves of the target QSOs are presented in
Figures \ref{f_lc1} and \ref{f_lc2},
and the parameters of the light curves
are listed in Table \ref{t_filterwave}. 
We found that all targets showed significant flux variations
during 7 years of observation, that is, 
a time span of $2-3.5$ years in the rest frame of the target QSOs.

\section{Examination of UV Color Variability}
\label{s_result}

\subsection{Flux-to-flux Plot and Linear Fit}
\label{s_ffplotfit}

\begin{figure}
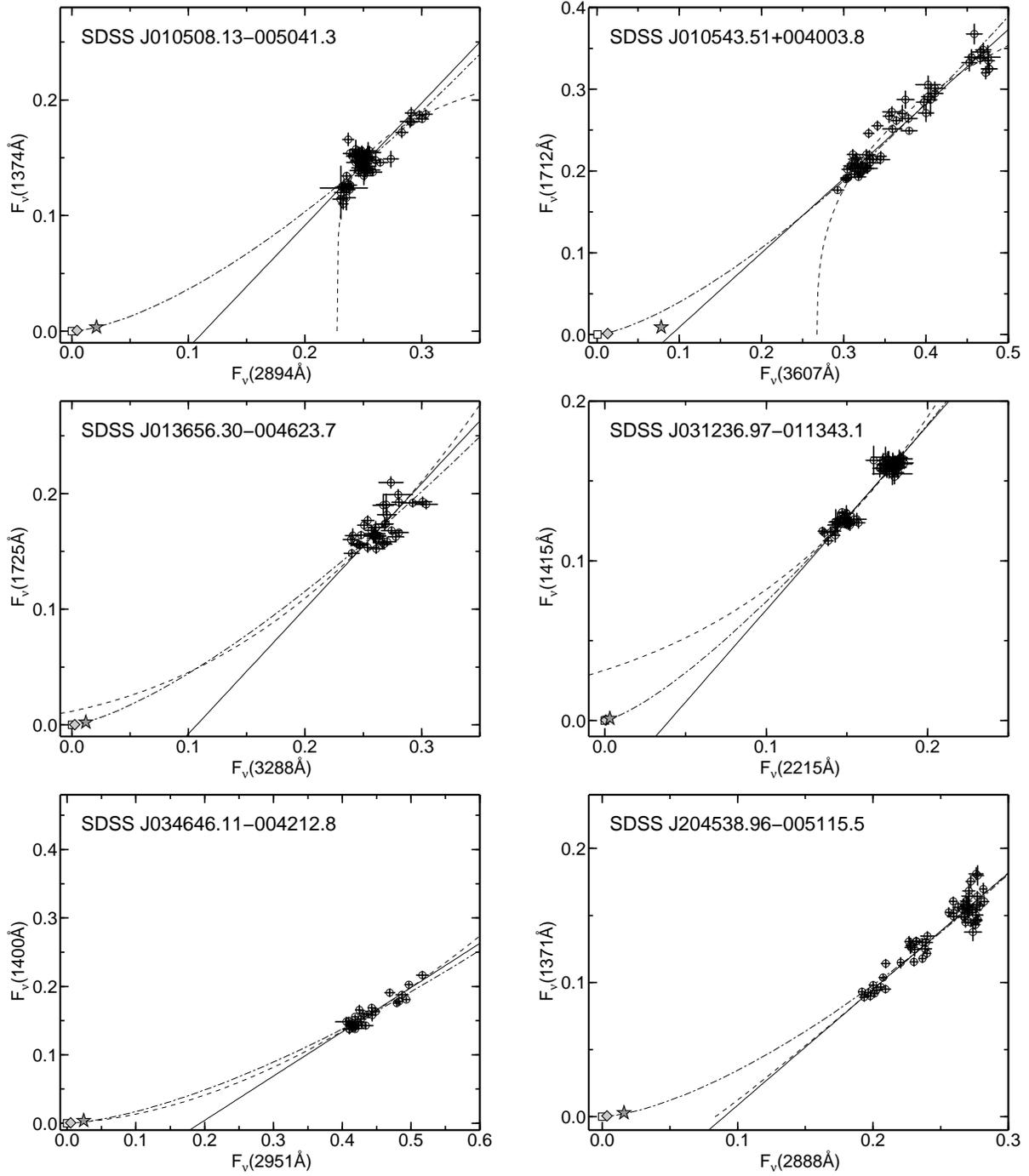

\plottwo{fig11.eps}{fig12.eps}

\plottwo{fig13.eps}{fig14.eps}

\plottwo{fig15.eps}{fig16.eps}

\caption{Flux-to-flux diagrams for SDSS J010508.13-005041.3, SDSS J010543.51+004003.8, 
SDSS J013656.30-004623.7, SDSS J031236.97-011343.1, SDSS J034646.11-004212.8, 
and SDSS J204538.96-005115.5. 
The unit of flux is mJy.
Open circles with error bars represent the multi-epoch flux data.
Square, diamond, and stellate symbols represent the host galaxy flux 
for the galaxy types E, Sbc, and Im, respectively. 
Thick, dashed, and dashed-dotted lines represent
the best-fit regression of a straight line, a power-law,
and a power-law whose power is derived from
the wavelength-dependent amplitude of variation of \citet{2004ApJ...601..692V}.  
\label{f_ff1}
}
\end{figure}

In order to examine the color variability of the UV continuum emission of QSOs 
with flux variation, we applied the flux-to-flux plot analysis
as applied to the color variability of optical continuum emission
of Seyfert galaxies in Paper I, which was originally proposed by
\citet{1981AcA....31..293C}. 
We plotted the flux data in two different bands observed on the same night
in the flux-to-flux diagram,
where the flux in longer wavelength is assigned to the $x$-axis
and the flux in shorter wavelength is assigned to the $y$-axis.
The flux-to-flux plots for all target QSOs are presented
in Figures \ref{f_ff1} and \ref{f_ff2}, and
the band pair and its effective wavelength in the rest frame
for each target are listed in Table \ref{t_filterwave}. 
The rest wavelength of the shorter-$\lambda$ filter is either $\sim1400$ \AA \, 
or $\sim1730$ \AA \,, 
while that of the longer-$\lambda$ is between $2200-3600$ \AA. 

As shown in Figures \ref{f_ff1} and \ref{f_ff2},
the rest-frame UV flux data in the two different bands
are distributed linearly for all targets,
and especially for those with large flux variation 
such a correlation seems very tight.
In order to examine the linear relationship between
the UV fluxes in the two different bands,
both straight-line fitting and power-law fitting to the data
in the flux-to-flux diagram were carried out
using the fitting code made by \citet{tomitaphd}
following the most generalized multivariate
least square process given by Jefferys (1980, 1981).
The fitting functions adopted for the straight-line fitting
and the power-law fitting are
$F_{\nu 1}=\alpha \times (F_{\nu_{2}}-F_{\nu_{2}0})$
and $F_{\nu 1}=\alpha \times (F_{\nu_{2}}-F_{\nu_{2}0})^\beta $, respectively,
because larger contamination of the host galaxy flux is expected
in the longer-wavelength flux ($F_{\nu_{2}}$)
than in the shorter-wavelength flux ($F_{\nu_{1}}$).
The best-fit parameter values and the reduced $\chi^2$ values
of the straight-line fitting and the power-law fitting
are listed in Table \ref{t_fffit}. 

As listed in Table \ref{t_fffit},
the residuals of the straight-line fitting
of the flux-to-flux plots in UV wavelengths
are so small that the reduced $\chi^2$ is near unity,
except for J2045$-$0051.
Indeed, according to the $\chi^2$-test for the straight-line fitting,
the linear relationship is not rejected for 6 of the 10 targets
(J0312$-$0113, J0346$-$0042, J2111$+$0024, J2119$+$0032, J2123$-$0050,
and J2134$+$0048) at a 1\% level of significance.
Although the $\chi^2$-test rejects the linear relationship
at a 1\% level of significance for the remaining four targets,
no clear curvature can be seen in their flux-to-flux plots.
Indeed, no significant improvement of $\chi^2$ values
applying the power-law fitting instead of the straight-line fitting
can be found for two of the four targets (J0136$-$0046 and J2045$-$0051).
The larger $\chi^2$ values of the straight-line fitting
of the four targets are probably caused by
another variable component of fluxes that is
not synchronized to the UV continuum emission,
or by underestimation of the photometric errors.
In fact, as will be described in Section \ref {s_effectsbp},
the reduced $\chi^2$ values of the straight-line fitting
are significantly decreased for three of the four targets
(J0105$-$0050, J0105$+$0040, and J2045$-$0051),
and the linear relationship is not rejected at a 1\% level of significance
for two targets (J0105$-$0050 and J0105$+$0040) 
when the variable flux component of the SBB is considered.
Based on these considerations,
we conclude that the UV flux of all of the target QSOs
in the two different bands shows a linear correlation
during the flux variations.

\subsection{Location of Host-Galaxy Component in Flux-to-flux Diagram}
\label{s_effectotagncont}

The x-intercept of the best-fit linear regression line,
$F_{\nu_{2}0}$, is positive for 9 of the 10 targets, 
as listed in Table \ref {t_fffit}
(with the exception of J2111$+$0024, for which
$F_{\nu_{2}0}$ is consistent to be zero within $2\sigma $ error). 
Hence, their rest-frame UV colors of the fluxes
observed in the two different bands become bluer as they become brighter.
However, as presented in the discussion in Paper I, 
if the sum of the non-variable flux components, such as that of the host galaxy, 
is located on the fainter extension of the best-fit regression lines
of the flux-to-flux plots,
the rest-frame UV color is almost constant during the flux variation.
In this section, we estimate the host galaxy fluxes of the target QSOs
and examine whether the UV colors of the targets 
show the trend of spectral hardening
or remain almost constant during the flux variations.

Since the host galaxies of the target QSOs are not
spatially resolved in the SDSS images,
we first estimated the stellar mass of the host galaxy
from the black hole mass estimated in Section \ref {s_targetselec}
and the Magorrian relation \citep{1998AJ....115.2285M},
a tight correlation between the central black hole mass
and the stellar mass of the bulge of the host galaxy. 
Next, the host galaxy fluxes in observing bands were converted
from the bulge mass using a mass-to-light ratio
and an SED of stellar population of the host galaxy. 
We assumed bulge luminosity as the whole luminosity of
the host galaxy 
because the QSOs at $z\lesssim 2$ are mostly hosted
by the bulge-dominated early-type galaxies
(e.g., Bahcall et al. 1997; Hutchings et al. 2002;
Dunlop et al. 2003; Zakamska et al. 2006) 

We used the Magorrian relation presented in \citet{2003ApJ...589L..21M},
\begin{equation}
\log M_{\rm BH}= 8.28 + 0.96 \times \log (M_{\rm bulge } -10.9),
\end{equation}
to estimate the stellar mass of the host galaxy from the black hole mass.
The scatter of the correlation was estimated as 0.21 dex by them.
Including the scatter of the black hole mass estimation
based on the scaling relation for \ion{C}{4} against
UV luminosity 0.32 dex presented in \citet{2006ApJ...641..689V} 
we estimated the scatter of the stellar mass of the host galaxy
as 0.38 dex, that is, a factor of 2.4 for $1\sigma $ error
and 5.8 for $2\sigma $ error.
Although the redshift evolution of the Magorrian relation
is still uncertain, \citet{2006ApJ...649..616P} suggested that
the $M_{\rm BH}/M_{\rm bulge}$ mass ratio
increases a factor of $\sim 4$ at $z>1.7$
and \citet{2010MNRAS.402.2453D} also showed that
it increases a factor of $\sim 7$ at $z=3$
compared to that of the local universe.
Therefore, the stellar mass, thus the luminosity of the host galaxy
of the targets estimated here, might be systematically overestimated
by a factor of $\sim 5$.

Contrary to the morphology, recent studies show that
the optical colors of AGN host galaxies at $z\lesssim 1.5$ 
are slightly bluer than those of quiescent early-type galaxies,
indicating a component of newly formed stellar population
(e.g., Jahnke, Kuhlbrodt, \& Wisotzki 2004b;
S\'{a}nchez et al. 2004; Nandra et al. 2007; Silverman et al. 2008).
In addition, at higher redshift,
\citet{Jahnke2004b} showed that
the rest-frame UV colors of host galaxies of QSOs at $1.8<z<2.75$
are bluer than that expected from an old stellar population with
a formation epoch at $z\sim 5$,
suggesting a young stellar population
of a few percent of the total mass. 
Schramm, Wisotzki, \& Jahnke (2008) showed that
the rest-frame $B-V$ colors of the host galaxies of three QSOs
at $z\sim 3$ are close to zero,
indicating a substantial contribution from young stars, 
and a stellar mass-to-light ratio below 1.

Since Kiuchi, Ohta, \& Akiyama (2009) demonstrated that
most optical SEDs of host galaxies of type-2 AGNs at $0.5<z<1.15$
follow composite SEDs of E to Sbc galaxies of the local universe,
and since \citet{Barger2003} showed that
the obs-frame $R-HK'$ colors of host galaxies of type-2 QSOs at $z=1-2$
are widely scattered between the K-corrected colors of
composite SEDs of local E and Im galaxies,
we adopt a local Sbc galaxy as a typical model
for estimating the mass-to-light ratio and the SED
of host galaxies of the target QSOs; 
we use local E and Im galaxies for the comparison.
We use mass-to-light ratios of 20, 2, and 0.5
for E, Sbc, and Im, respectively
\citep{1976ApJ...204..668F, 2001ApJ...550..212B},
and calculate the fluxes of the host galaxies
in the rest-frame wavelength coverage of the observing filters
using an UV-optical spectrum template of
nearby galaxies for E, Sbc, and Im presented
in \citet{2008ApJ...676..286A}. 

Figures \ref{f_ff1} and \ref{f_ff2} present
the location of the host galaxy flux of the target QSOs
in the flux-to-flux diagrams.
The estimated host galaxy fluxes, as described, 
are listed in Table \ref{t_host}.
Figures \ref{f_ff1} and \ref{f_ff2} clearly show that
the location of the host galaxy flux is systematically
on the left side of the best-fit regression line.
For a typical case of the Sbc-type galaxy,
the host galaxy fluxes of 9 of the 10 targets
are different from the fainter
extension of the best-fit regression line
by more than $2\sigma $ error of the host galaxy fluxes,
or a factor of 5.8.
The only target for which the host galaxy flux is located
on the extension of the best-fit regression line
within $2\sigma $ error, J2115$+$0024, is in fact 
$F_{\nu _{2}0}$ consistent to be zero, 
and the host galaxy flux is estimated to be considerably small.
Instead, if we assume the Im-type galaxy for the host galaxy
to model recent star-forming activity, 
as reported for high-redshift QSOs by Schramm et al. (2008),
the host-galaxy fluxes of 5 of the 10 targets
(J0105$-$0050, J0105$+$0040, J2045$-$0051, J2111$+$0024, and J2134$+$0048)
are located on the extension of the best-fit regression line
within $2\sigma $ error.
However, if the estimated black hole mass
using the Magorrian relation at the local universe
is systematically smaller by a factor of $\sim 5$
caused by the redshift evolution,
the location of the host galaxy flux
is on the left side of the best-fit regression line
for four of those five targets
(except the only target, again J2115$+$0024).

We conclude that most of the target QSOs show
spectral hardening in UV wavelengths, that is,
the UV continuum emission becomes bluer as it becomes brighter,
in spite of maintaining the linear correlations between the QSOs' two-band flux data.
These results suggest that the UV continuum emission of QSOs
usually shows spectral hardening, which is consistent with
the results reported by \citet{2004ApJ...601..692V}
and \citet{2005ApJ...633..638W}
from the colors of the differential fluxes
or the composite differential spectrum
of two-epoch observations for SDSS QSOs.

\subsection{Effect of Small Blue Bump}
\label{s_effectsbp}
The best-fit parameters and the reduced $\chi ^2$ values 
of the straight-line fitting are listed in Table \ref{t_fffit}.
As described in Section \ref{s_ffplotfit},
the reduced $\chi ^2$ value is so large
that the linear relationship is rejected 
at a 1\% level of significance for 4 of the 10 targets 
(J0105$-$0050, J0105$+$0040, J0136$-$0046, and J2045$-$0051).
These larger $\chi ^2$ values suggest contamination from 
another variable component of fluxes
that is not synchronized to the UV continuum emission.

The SBB consisting of \ion {Fe}{2} emission lines
and Balmer continuum emission,
and the \ion {Mg}{2} $\lambda 2798$ emission line,
which more or less contaminate our observing bands
of the longer-wavelengths,
are thought to be the most promising sources
for such additional variable components.
They are considered to originate in the broad emission-line region (BLR),
and 
moreover, reverberation mapping observations
have found the lag of the SBB flux variation 
behind that of the UV continuum emission
for NGC 5548 \citep{1993ApJ...404..576M} 
and the lag of the \ion {Mg}{2} emission-line flux variation 
for NGC 4151 \citep{metzroth2006}.
Since the size of the BLR of the target QSOs
is estimated to be more than 200 light days
for H$\beta $ emission line by the luminosity scaling relation
\citep{2005ApJ...629...61K},
the flux variation of the SBB emission and the \ion {Mg}{2} emission line
is not at all synchronized with that of the continuum emission 
and would provide a scatter around the linear correlation
between the fluxes of the UV continuum emission in the two different bands.
We estimated the scatter of the linear correlation
in the flux-to-flux plots caused by 
the contaminations from the SBB and \ion{Mg}{2} emissions 
and re-examined the linear relationship of
the UV continuum emission in the two different bands.

We first fit the target spectrum obtained from the SDSS database
by a power-law model with the spectral windows
of $\lambda=1440-1470$\AA $\,, 1685-1695$\AA, $2200-2230$\AA, and $4000-4020\AA$,
where the spectrum is free from any strong emission lines.
The synthesized flux of the spectrum in the photometric band
was estimated by convolving the spectrum with the filter transmission curve,
and then, the flux of the spectrum was recalibrated by
scaling the synthesized flux by the photometric data
at the observing date of the spectroscopy
that was estimated by interpolating the light curve.
The contaminated flux of the SBB and \ion{Mg}{2} emissions
in the broad-band filter was estimated by subtracting the synthesized flux
of the best-fit power-law spectrum from that of the original spectrum.
The contaminated fluxes of the SBB and \ion{Mg}{2} emissions
in the longer wavelength filter are listed in Table \ref{t_sbb}. 
Their uncertainties were derived from the error of the power-law fitting.

Next, the $1\sigma $ scatter of the flux variation
of the SBB and \ion{Mg}{2} emission-line component,
$\sigma _{SBB}$, was estimated as $f_{SBB}\times \sigma_{c}/\bar{f_{c}}$, 
where $f_{SBB}$, $\bar{f_{c}}$, and $\sigma_c$ are
the contaminated flux of the SBB and \ion{Mg}{2} emissions,
the average flux, and the $1\sigma $ scatter
over the broad-band light curve, respectively.
The $\sigma _{SBB}$ is also listed in Table \ref{t_sbb}.

Finally, assuming that the flux variation of
the SBB and \ion{Mg}{2} emission line
is totally uncorrelated with that of the continuum emission in time,
we recalculated the reduced $\chi^2$ of
the straight-line fitting for the flux-to-flux plots
by replacing $\sigma _{obs}$ by $\sqrt{\sigma_{obs}^2+\sigma_{SBB}^2}$,
where $\sigma _{obs}$ is the observational error of the photometry.
The new reduced $\chi^2$ values of the straight-line fitting
in which the scatter caused by the contaminations of
the SBB and \ion{Mg}{2} emissions
is considered, are listed in Table \ref{t_sbb}. 

We found that
the reduced $\chi^2$ values of the straight-line fitting
were significantly decreased for three targets
(J0105$-$0050, J0105$+$0040, and J2045$-$0051)
out of the four targets,
for which the linear relationship was rejected
by the $\chi ^2$ analysis in Section \ref {s_ffplotfit},
and the linear relationship is now not rejected
at a 1\% level of significance
for two targets (J0105$-$0050 and J0105$+$0040).
These results strongly indicate that
a part of the scatter of the flux-to-flux plot from 
the best-fit regression line is caused by
the contaminations from the SBB and \ion {Mg}{2} emission line,
and the UV continuum emissions in two different bands
of the target QSOs show a tight linear correlation
during the flux variations.

\section{Discussion}
\label{s_discuss}

\subsection{Comparison with Previous Studies}
\label{s_compprevstudy}

\subsubsection{Vanden Berk et al. (2004)}
\label{s_vandenberk}

\citet{2004ApJ...601..692V}
examined the relationship between variability amplitude
and the rest-frame wavelength of the UV-optical region
from SDSS two-epoch multi-band observations
of about 25,000 QSOs,
and found a larger amplitude of variation
at shorter wavelengths in the UV region of $\lambda < 4000$ \AA.
In this section, the tracks of the flux-to-flux plot of the target QSOs
obtained from the photometric monitoring data,
which show the hardening trend in the UV region, as presented in the previous sections,
are compared with the hardening trend of the amplitude of variation 
presented in \citet{2004ApJ...601..692V}.

Assuming that the wavelength-dependent amplitude of variation 
always holds for the flux variation at all times for individual QSOs,
a power-law function of $F_{\nu {1}}=\alpha \times F_{\nu {2}}^\beta $ in a flux-to-flux plot
is obtained, where $\beta =v(\lambda_1)/v(\lambda_2) $ is
derived from the wavelength-dependent amplitude of variation $v(\lambda )$,
presented in Equation 11 of \citet{2004ApJ...601..692V}.
Then, we fit the power-law function,
$F_{\nu {1}}=\alpha_{2} \times F_{\nu _{2}}^{\beta_{2}}$,
to the flux-to-flux plot data of the target QSOs,
fixing the parameter $\beta _{2}$ as $\beta_{2}=v(\lambda_1)/v(\lambda_2)$, 
where $\lambda_1$ and $\lambda_2 (\lambda_1 <\lambda_2)$ are
the effective wavelengths of the observing filters in the rest frame. 

Figures \ref{f_ff1} and \ref{f_ff2} present the fitting of the power-law
function to the flux-to-flux plots, and the best-fit parameters
and the reduced $\chi ^2$ values are listed in Table \ref{t_vanden}.
The reduced $\chi^2$ value of the power-law fit is 
comparable to that of the straight-line fit for most of the target QSOs.
The slope of the power-law fit seems to be
slightly gentler than that of the flux-to-flux plots for J2123$-$0050,  
and in fact, the reduced $\chi ^2$ of the power-law fit
is slightly larger than that of the straight-line fit
even if it is still as small as $\sim 1$.
When we refit the power-law function, 
$F_{\nu {1}}=\alpha_{2} \times F_{\nu _{2}}^{\beta_{2}}$, 
with both $\alpha _{2}$ and $\beta _{2}$ freed,
$\beta _{2}=1.908\pm 0.152$ is derived from the new power-law fit
with the reduced $\chi ^2=0.695$.
The $\beta _{2}$ value is slightly different from $v(\lambda_1)/v(\lambda_2) $ 
but would be still consistent with \citet{2004ApJ...601..692V}
within the fitting error and the diversity of QSO properties, 
and the reduced $\chi ^2$ becomes comparable to that of the straight-line fit.
From these results, we conclude that
our results of the hardening trend during the flux variation
obtained from the flux-to-flux plots in the rest-frame UV wavelengths
are consistent with the hardening trend of
the variability amplitude in the UV region presented in \citet{2004ApJ...601..692V}.

\subsubsection{\citet{1996A&A...312...55P}}
\label{s_paltaniwalter}

\citet{1996A&A...312...55P} observed the UV spectra of 15 nearby AGNs 
at different epochs using the IUE satellite.
Applying PCA to the multi-epoch spectra and the decomposition 
to a variable and a non-variable components, 
they explained the UV variation in almost all their targets 
as the sum of a variable component of a constant power-law spectral shape
and a non-variable component of the SBB,
which means that the spectral shape of the UV continuum emission
is almost constant during the flux variation.
This finding is not consistent with our results
that the UV continuum emission becomes bluer
as it becomes brighter for most of the target QSOs.
In fact,  \citet{1996A&A...312...55P} did not
estimate the SBB fluxes of the target AGNs
by fitting a spectral model of a power-law continuum plux the SBB 
to their UV spectrum in order to examine the consistency 
of that method with the conclusion from the decomposition to 
a variable and a non-variable components.
Moreover, as discussed in Section \ref{s_effectsbp},
the SBB is thought to originate in the BLR 
and its flux was found to be variable.
It is suspected that a sufficient amount of the SBB flux is
constant, which explains the constant spectral shape
of the UV continuum emission during flux variation.

On the other hand, Paltani \& Walter's finding that 
one eigenvalue dominated all the others in the PCA of the UV flux variation 
is consistent with the linear relationship of the fluxes
in the two different bands found in the flux-to-flux plots of our target QSOs.
The models of the accretion disk predict that
the spectral shape of the UV-optical continuum emission
depends on the mass accretion rate and the black hole mass; 
thus, the spectral properties of flux variation
would also depend on those AGN properties.
Therefore, while it is beyond the scope of this paper,
it is very interesting to re-examine
the multi-epoch UV spectra of nearby AGNs of \citet{1996A&A...312...55P}
using the flux-to-flux plot analysis with accurate estimation of
non-variable components such as the host galaxy 
and also to discuss the luminosity or the black hole mass
dependency of the spectral variability of the UV continuum emission.

\subsection{Comparison with Accretion Disk Model}
\label{s_accratemodel}

We showed that the spectral shape of the optical continuum emission 
is almost constant during the flux variation for 11 Seyfert galaxies in Paper I,
which suggests that the radial temperature profile of 
an optical emitting region of an accretion disk
does not change during the flux variation.
On the other hand,
we here find that the UV continuum emission becomes bluer as it brightens
for most of the target QSOs, while the UV fluxes in the two different bands
show a tight linear correlation.
This suggests that
the temperature structure of the UV emitting region of an accretion disk
shows systematic change.

When the mass accretion rate changes in a
standard accretion disk model \citep{1973A&A....24..337S},
the spectral shape of the continuum emission from the accretion disk
changes in UV wavelengths
while it remains almost constant in optical wavelengths for AGNs 
whose central black hole mass is larger than $\sim 10^{7-8} M_{\odot}$ 
(e.g., Kawaguchi, Shimura, \& Mineshige 2001).
That is because the variation of the mass accretion rate 
changes the maximum temperature of the accretion disk,
which affects the spectral shape of UV continuum emission,
while it does not change the radial profile of outer accretion disk
of $T(r)\propto r^{-3/4}$ where the optical continuum emission originates, 
which would result in an almost constant optical continuum spectral shape.

\citet{2006ApJ...642...87P} presented that 
a standard accretion disk model in which
the mass accretion rate changed from one epoch to the next
could be successfully fitted to the composite differential spectrum
of two epochs of observations for hundreds of SDSS QSOs,
which is bluer than the composite spectrum of QSOs
indicating the spectral hardening of UV continuum emission
\citep{2005ApJ...633..638W}.
However, the composite spectrum represents
the average characteristics of QSOs; 
moreover, the differential spectrum corresponds 
only to the slope of the linear relationship between fluxes
in two different bands in the flux-to-flux plots.
Here, we fit a standard accretion disk model
with a varying mass accretion rate
to the flux-to-flux plots for individual target QSOs,
and examine whether the spectral variation of the UV continuum emission
during flux variations at various epochs
can be described by the standard accretion disk model
with various mass accretion rates and a constant black hole mass.

The inner radius of the standard accretion disk model
is set as $R_{in}=3Rs$, where $Rs$ is the Schwarzschild radius.
The outer radius is set as $R_{out}=1000Rs$,
but it is not important for the UV spectrum.
A face-on view is assumed to calculate the flux from the accretion disk.
Then, a trajectory of the accretion disk model in the flux-to-flux diagram
can be traced by changing the mass accretion rate
with a constant black hole mass.
We fit the trajectory of the accretion disk model
to the flux-to-flux plots of the individual target QSOs 
with a free parameter of black hole mass.
Since the observed flux includes the contaminating fluxes of 
the host galaxy, the SBB and \ion{Mg}{2} emission line
in addition to the continuum emission,
the Sbc-type host galaxy flux estimated in Section \ref{s_effectotagncont},
and the SBB and \ion{Mg}{2} contaminating flux
estimated in Section \ref{s_effectsbp}
are added to the trajectory of the accretion disk model
before fitting.

\begin{figure}
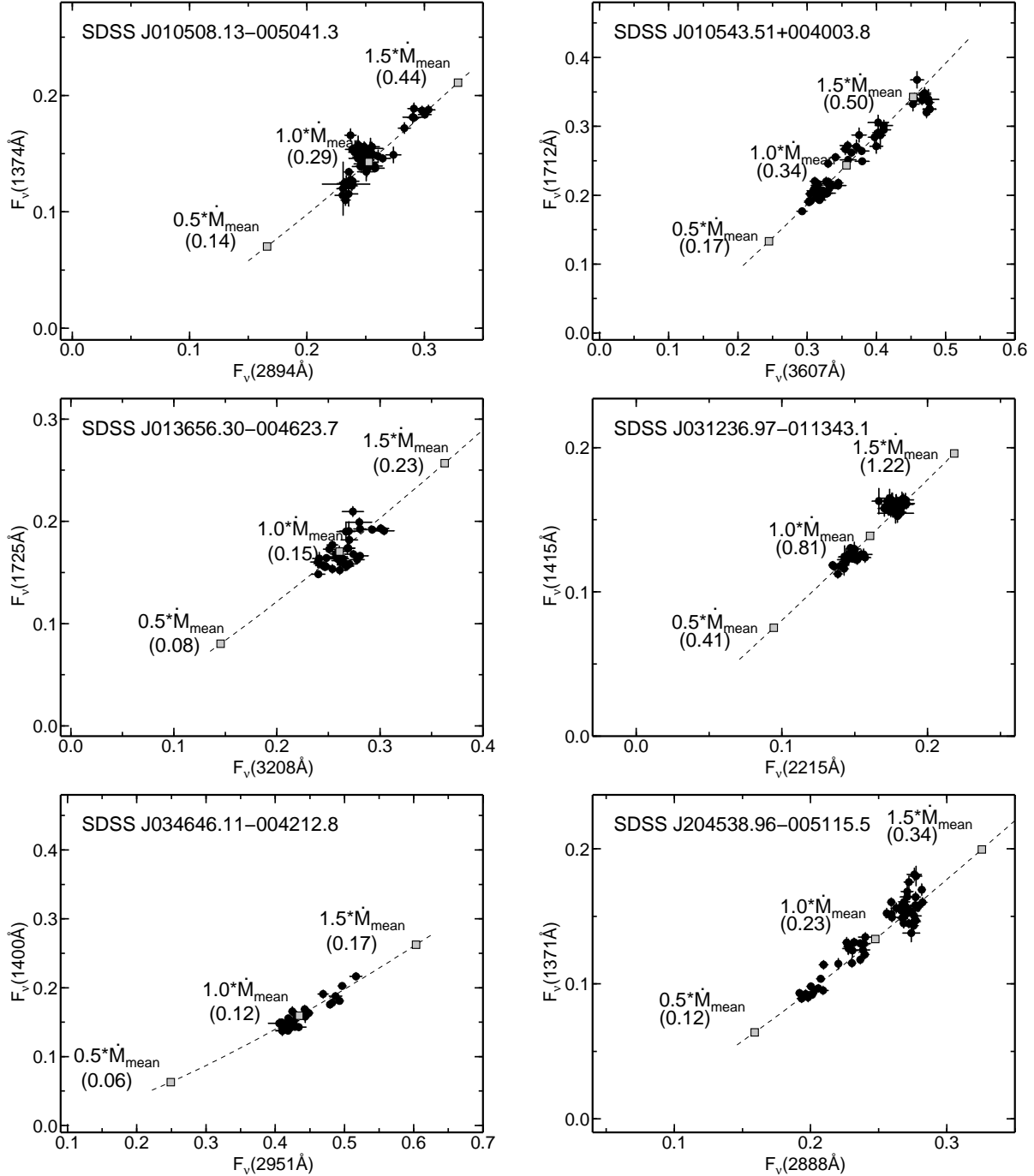

\plottwo{fig21.eps}{fig22.eps}

\plottwo{fig23.eps}{fig24.eps}

\plottwo{fig25.eps}{fig26.eps}

\caption{
Multi-epoch flux data compared with
the standard accretion disk model
with varied mass accretion rates and a constant black hole mass
for SDSS J010508.13-005041.3, SDSS J010543.51+004003.8, 
SDSS J013656.30-004623.7, SDSS J031236.97-011343.1, SDSS J034646.11-004212.8, 
and SDSS J204538.96-005115.5. 
Open circles with error bars represent the multi-epoch flux data,
and a dashed line represents the curve
in the flux-to-flux diagram of the best-fit model.
Three squares are marked on the best-fit model curve
corresponding to 0.5, 1.0, and 1.5 times
the mean mass accretion rate
that will produce the fluxes averaged over the light curves.
The Eddington ratios of the best-fit model
are provided in parentheses next to the squares.
\label{f_ffsim1}
}
\end{figure}

\begin{figure}

\plotone{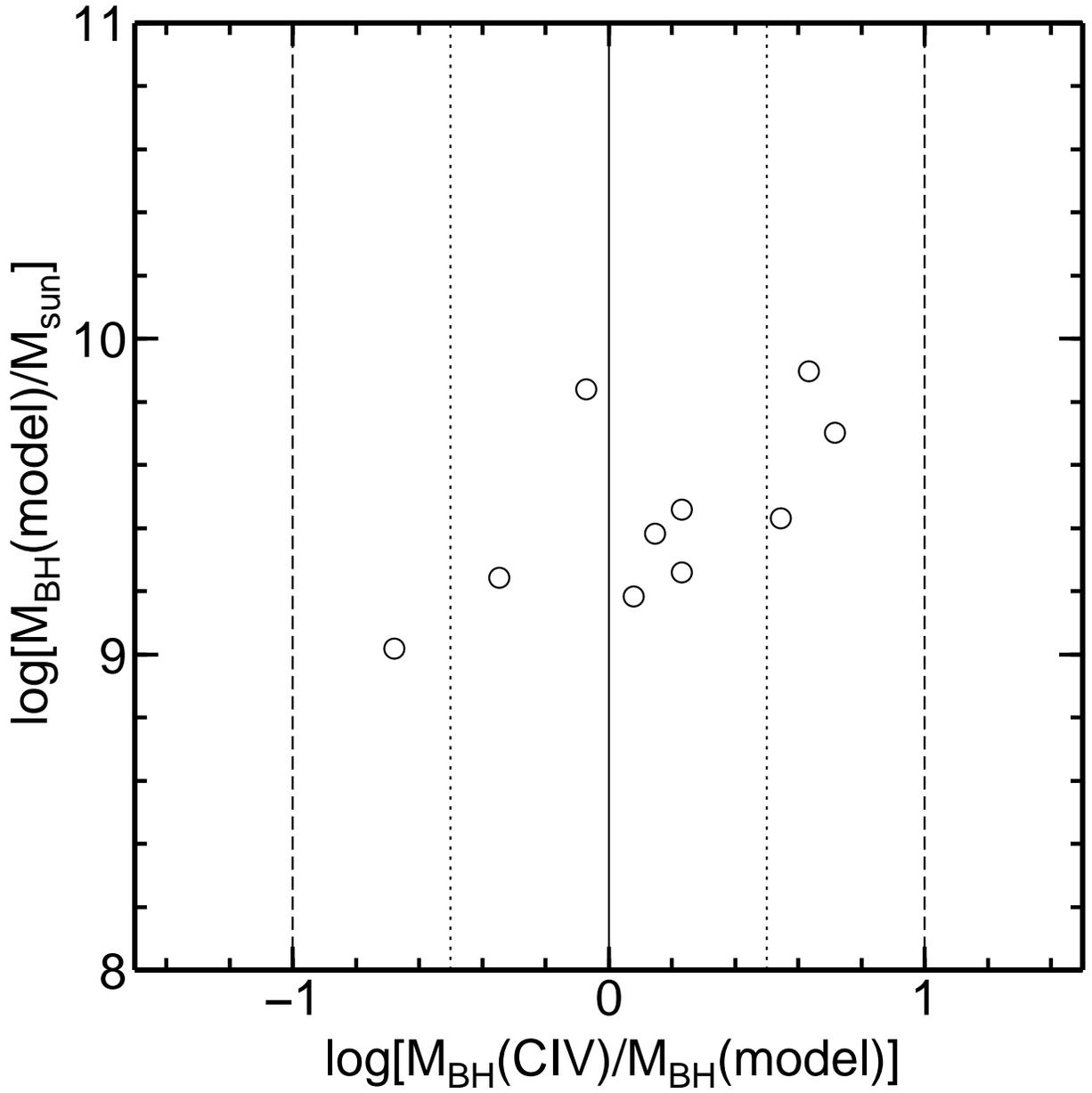}

\caption{The ratio of the black hole mass estimated from
the luminosity scaling relation and the emission-line width
to that of the best-fit standard accretion disk model
applied to the flux-to-flux plots.
\label{f_bhmass}
}
\end{figure}

Figures \ref{f_ffsim1} and \ref{f_ffsim2} present
the best-fit accretion disk model for the flux-to-flux plots
of the target QSOs, and the best-fit black hole mass and
the reduced $\chi ^2$ are listed in Table \ref{t_simstat}.
The error of the black hole mass in the table is estimated from
the statistical uncertainty of $\chi^2$-fitting of the model.
As shown in Figures \ref{f_ffsim1} and \ref{f_ffsim2},
the flux-to-flux plots can be fitted well
by the standard accretion disk model, changing the mass accretion rate
with a constant black hole mass.
Indeed, the reduced $\chi^2$ values of the standard accretion disk model
are comparable to those of the straight-line fitting
for 9 of the 10 target QSOs, and even for the exception, 
J2123$-$0050, the reduced $\chi ^2$ is as small as $\sim 1$.
%
In addition, as shown in Figure \ref{f_bhmass},
the best-fit black hole mass of the standard accretion disk model
is consistent with the black hole mass estimated from the emission-line width
and the luminosity scaling relation in Section \ref{s_targetselec},
within 0.76 dex for $2\sigma $ error of
the latter black hole mass estimation.
We also calculate the bolometric luminosity of
the best-fit standard accretion disk model whose mass accretion rate
corresponds to the average flux of the light curve.
The ratio of the bolometric luminosity
to the specific luminosity in UV,
the bolometric correction $L_{\rm bol}/\lambda L_{\lambda }(3000{\rm \AA})$,
is listed in Table \ref{t_simstat}.
We find that the bolometric correction
is consistent with those obtained for
hundreds of type 1 QSOs from \citet{2006ApJS..166..470R},
which are distributed between 4 to 9 and whose average value is 5.62.
The black hole mass and the luminosity of the best-fit
standard accretion disk model fitted to the flux-to-flux plots
in UV wavebands are reasonable, and it could be a new technique for
estimating a black hole mass of luminous QSOs although the systematic
uncertainties should be examined.

As presented in Paper I, the optical color of
the best-fit regression line in the flux-to-flux diagram
can be regarded as that of the AGN optical continuum,
because the optical spectral shape remains almost constant
during the flux variation and the colors derived from
the flux-to-flux plot analysis of nearby Seyfert galaxies
were consistent with those of the standard accretion disk model,
$\alpha _{\nu}=1/3$.
Although it has long been known that the standard accretion disk model
apparently shows a contradiction with observations
that the composite spectrum of QSOs is redder than that of
the standard accretion disk model
\citep{1991ApJ...373..465F,2001AJ....122..549V},
the detailed analyses of 
the spectral shape of the UV-optical continuum emission of AGNs
using flux variation presented here and in Paper I
strongly support the standard accretion disk.
A possible explanation for the composite spectra being redder is 
that host galaxy light contaminates the spectra preferentially 
at longer wavelengths.
These results are consistent with the recent analyses
of the spectral shape of the optical to near-infrared continuum emission, 
which comes from the accretion disk
by flux variations \citep{2006ApJ...652L..13T}
and polarizations \citep{2008Natur.454..492K}.

However, it has a critical difficulty 
of considering the variation
of global accretion rate
of a standard accretion disk
as a primary source of flux variation of QSOs.
The timescale of changes of the global mass accretion rate
is believed to correspond to the viscous timescale
(e.g., Pringle 1981; Frank et al. 2002),
which is about $\sim 10^6$ years for our target QSOs
and much longer than the timescale of the flux variation we observed.
Clearly,
further study of the flux variation mechanism is desired,
which can explain the observed properties
of spectral variations
of the UV-optical continuum emission of AGNs,
suggesting a variation of the characteristic
temperature of an accretion disk.

\section{Summary}
\label{s_summary}

We examined the spectral variability of UV continuum emission of AGNs
based on the multi-epoch photometric data of 10 SDSS QSOs
in the Stripe 82.
The target redshift was $z=1.05,1.54,1.71, 2.35\pm0.05$ 
in which strong broad lines such as Ly$\alpha$ and \ion{C}{4}
do not contaminate SDSS filters.
The target luminosity was about $\lambda L_{\lambda }(3000{\rm \AA})\sim 10^{46-47}$ erg s$^{-1}$,
the black hole mass was estimated to be about $10^{9-10}M_{\odot}$
based on the luminosity scaling relations
for \ion{C}{4} or \ion{Mg}{2} emission lines and their line width,
and the mean Eddington ratio was between $0.1-1.2$.
All target QSOs showed significant flux variation
within 7 years of observation, or $2-3.5$ years in the rest-frame.

We plotted the flux data in two different bands 
(shorter-$\lambda_{rest}\sim 1400$ \AA \, or $\sim 1730$ \AA, 
longer-$\lambda_{rest}\sim2200$ \AA $- 3600$ \AA) 
observed on the same night in the flux-to-flux diagram 
and found a strong linear correlation 
between the fluxes in the two different bands 
during the flux variations for all target QSOs.
The linear relationship was not rejected for 6 of the 10 targets 
at a 1$\%$ level of significance based on the $\chi^2$-test
for straight-line fitting.
Even for the remaining four targets,
no clear curvature could be seen in their flux-to-flux plots;
no significant improvement of $\chi^2$ values by the power-law fitting 
than those of the straight-line fitting could be found
for two of the four targets, 
and the reduced $\chi^2$ values of the straight-line fitting
were significantly decreased for three of the four targets
when the contamination of the SBB was considered.

We then examined the location of the host galaxy flux in the flux-to-flux diagram.
The stellar mass of the host galaxy
was estimated from the black hole mass
on the basis of the Magorrian relation \citep{2003ApJ...589L..21M}; 
then, the host galaxy flux was estimated from the stellar mass
using the mass-to-light ratio and the SED of a local Sbc galaxy
\citep{2001ApJ...550..212B,2008ApJ...676..286A}.
We found that the location of the host galaxy flux was systematically on the left side 
of the best-fit regression line for 9 of the 10 targets; 
that is, 
the UV continuum emission becomes bluer as it brightens
for most of the target QSOs
in spite of holding their linear correlations between the two-band flux data.  
This trend of spectral hardening in UV wavelengths
is consistent with the findings of \citet{2004ApJ...601..692V}.
Indeed, we found that the flux-to-flux plots of the target QSOs 
could be fitted well by a power-law function assuming
the wavelength-dependent variability amplitude of QSOs.

We presented in Paper I that 
the spectral shape of the optical continuum emission of AGNs
remains almost constant and that its color is consistent with that of
the standard accretion disk model, $\alpha _{\nu }=1/3$.
Then, in the present paper, we fit a standard accretion disk model
with a changing mass accretion rate and a constant black hole mass
to the UV flux-to-flux plots for the individual target QSOs,
and found that the spectral variation of the UV continuum emission
during flux variations at various epochs
could be described by the standard accretion disk model
with varied mass accretion rates.
The black hole mass and the bolometric luminosity
from the best-fit standard accretion disk model
were reasonable.
Although it has long been known that 
the composite spectrum of QSOs is redder than that of
the standard accretion disk model,
the detailed analyses of 
the spectral shape of the UV-optical continuum emission of AGNs
using flux variation strongly support the standard accretion disk model.
However, 
the variation timescale of the global mass accretion rate
was considered to be too large
to explain the flux variations within a few years,
and further study of the flux variation mechanism is desired.

\acknowledgments

We thank Toshihiro Kawaguchi for useful discussion and comments. 
This research has been partly supported by the Grant-in-Aids of Scientific Research (10041110, 
 10304014, 12640233, 14047206, 14253001, and 14540223) and COE Research (07CE2002) 
of the Ministry of Education, Science, Culture and Sports of Japan.
Tomoki Morokuma has been supported by the JSPS (Japan Society for the Promotion of Science)
Research Fellowship for Young Scientists.

\clearpage

\clearpage

\begin{figure}
\plottwo{fig7.eps}{fig8.eps}

\plottwo{fig9.eps}{fig10.eps}

\caption{Same as in Figure \ref{f_lc1}, but for SDSS J211157.77+002457.5, SDSS J211908.43+003246.3, 
SDSS J212329.46-005052.9, and SDSS J213422.25+004850.3. 
\label{f_lc2}
}
\end{figure}

\clearpage

\clearpage

\begin{figure}

\plottwo{fig17.eps}{fig18.eps}

\plottwo{fig19.eps}{fig20.eps}

\caption{Same as in Figure \ref{f_ff1}, but for SDSS J211157.77+002457.5, SDSS J211908.43+003246.3, 
SDSS J212329.46-005052.9, and SDSS J213422.25+004850.3. 
\label{f_ff2}
}
\end{figure}

\clearpage

\clearpage

\begin{figure}

\plottwo{fig27.eps}{fig28.eps}

\plottwo{fig29.eps}{fig30.eps}

\caption{Same as in Figure \ref{f_ffsim1}, but for SDSS J211157.77+002457.5, SDSS J211908.43+003246.3, 
SDSS J212329.46-005052.9, and SDSS J213422.25+004850.3. 
\label{f_ffsim2}
}
\end{figure}

\clearpage

\clearpage

\begin{deluxetable}{lcccccccccc}
\tabletypesize{\footnotesize}
\rotate
\tablecaption{Light Curve of Target QSOs\label{t_filterwave}}
\tablewidth{0pt}
\tablehead{
\colhead{Object} & \colhead{Redshift} & \colhead{Filter$_1$} & \colhead{Filter$_2$} & 
\colhead{Rest $\lambda_1$\tablenotemark{a}[\AA]} & \colhead{Rest $\lambda_2$\tablenotemark{a}[\AA]}
& \colhead{Number of Data} &  \colhead{$\bar{F}_{\nu1}$\tablenotemark{b}} & \colhead{$\sigma_{\nu1}$\tablenotemark{c}} & \colhead{$\bar{F}_{\nu2}$\tablenotemark{b}} & \colhead{$\sigma_{\nu2}$\tablenotemark{c}} \\
}

\startdata
SDSS J010508.13$-$005041.3 & 1.585 & u & i & 1374 & 2894 & 68 & 0.146 & 0.017 & 0.251 & 0.017 \\
SDSS J010543.51$+$004003.8 & 1.074 & u & i & 1712 & 3607 & 57 & 0.238 & 0.056 & 0.361 & 0.059 \\
SDSS J013656.30$-$004623.7 & 1.716 & g & z & 1725 & 3288 & 39 & 0.169 & 0.015 & 0.263 & 0.015 \\
SDSS J031236.97$-$011343.1 & 2.311 & g & i & 1415 & 2215 & 67 & 0.140 & 0.019 & 0.161 & 0.016 \\
SDSS J034646.11$-$004212.8 & 1.535 & u & i & 1400 & 2951 & 33 & 0.156 & 0.020 & 0.436 & 0.030 \\
SDSS J204538.96$-$005115.5 & 1.590 & u & i & 1371& 2888 & 57 & 0.133 & 0.026 & 0.246 & 0.029 \\
SDSS J211157.77$+$002457.5 & 2.325 & g & i & 1409 & 2250 & 52 & 0.216 & 0.012 & 0.245 & 0.012 \\
SDSS J211908.43$+$003246.3 & 2.327 & g & i & 1408 & 2249 & 52 & 0.106 & 0.008 & 0.136 & 0.007 \\
SDSS J212329.46$-$005052.9 & 2.261 & g & i & 1437 & 2294 & 62 & 0.808 & 0.058 & 1.064 & 0.043 \\
SDSS J213422.25$+$004850.3 & 2.336 & g & i & 1405 & 2243 & 64 & 0.087 & 0.011 & 0.100 & 0.010 \\
\enddata
\tablenotetext{a}{The effective wavelength of the filter in the rest frame of the target QSO.}
\tablenotetext{b}{The average flux over the light curve in mJy.}
\tablenotetext{c}{The standard deviation of flux over the light curve in mJy.}
\end{deluxetable}

\clearpage

\begin{deluxetable}{lcccccccc}
\rotate
\tabletypesize{\tiny}
\tablecaption{Characteristics of Target QSOs \label{t_objbasicparm}}
\tablewidth{0pt}
\tablehead{
\colhead{Object} & \colhead{Radio\tablenotemark{a}} & \colhead{$\log(\lambda L_{\lambda}(1350\AA)$\tablenotemark{b}} & 
\colhead{$\log(\lambda L_{\lambda}(3000\AA)$\tablenotemark{b}} & \colhead{$\sigma($C IV$)$\tablenotemark{c}} & \colhead{$\sigma($Mg II$)$\tablenotemark{c}} & 
\colhead{$\log(M_{BH}($C IV$))$\tablenotemark{d}} & \colhead{$\log(M_{BH}($Mg II$))$\tablenotemark{d}} & \colhead{$L_{bol}/L_{Edd}$\tablenotemark{e}}  \\
 & & \colhead{[ergs s$^{-1}$]} & \colhead{[ergs s$^{-1}$]} & \colhead{[km s$^{-1}$]}& \colhead{[km s$^{-1}$]} & \colhead{[$M_{\odot}]$} & \colhead{[$M_{\odot}]$} & \\
}

\startdata
J0105$-$0050 & $\times$     & 46.30 & 46.16 & \, 2027 & 1720 & \, 9.29 & 9.21 & 0.33\\
J0105$+$0040 & $\times$     & ---   & 45.94 &     --- & 1848 &     --- & 9.19 & 0.27\\
J0136$-$0046 & $\times$     & 46.31 & 46.26 &   *4465 & 1340 &   *9.93 & 8.99 & 0.74\\
J0312$-$0113 & $\times$     & 46.47 & 46.40 & \, 1498 & ---  & \, 9.10 & ---  & 0.94\\
J0346$-$0042 & not covered  & 46.32 & 46.35 &   *1358 & 1746 &   *8.92 & 9.31 & 0.49 \\
J2045$-$0051 & not covered  & 46.26 & 46.16 & \, 2066 & 1622 & \, 9.20 & 9.09 & 0.34\\
J2111$+$0024 & $\times$     & 46.70 & 46.56 & \, 2392 & ---  & \, 9.64 & ---  & 0.37 \\
J2119$+$0032 & $\times$     & 46.37 & 46.34 & \, 1323 & ---  & \, 8.94 & ---  & 1.13\\
J2123$-$0050 & $\times$     & 47.24 & 47.20 & \, 2541 & ---  & \, 9.99 & ---  & 0.78\\
J2134$+$0048 & $\times$     & 46.24 & 46.17 &   *3731 & ---  &   *9.74 & ---  & 0.12\\
\enddata
\tablenotetext{a}{The results of VLA FIRST survey. Cross indicates not detected in the survey.}
\tablenotetext{b}{The luminosities at the epoch when the spectroscopic observations were done, computed by temporal and wavelength interpolation (or extrapolation) of the lightcurve shown in Figures \ref{f_lc1} and \ref{f_lc2}.}
\tablenotetext{c}{Standard deviation of the line from the archival SDSS spectroscopic data. 
Asterisk indicates that there is a strong feature of absorption in the line.}
\tablenotetext{d}{The central black-hole mass estimated based on single-epoch spectroscopy. Asterisk indicates computed from the line width showing strong absorption feature.}
\tablenotetext{e}{The Eddington ratio of the luminosity.}
\end{deluxetable}

\clearpage

\begin{deluxetable}{lcccccccc}
\tabletypesize{\tiny}
\tablecaption{Straight-line and Power-law Fits to Flux-to-flux Plots\label{t_fffit}}
\tablewidth{0pt}
\tablehead{
\colhead{Object} & \multicolumn{3}{c}{Straight-line Fit} & & \multicolumn{4}{c}{Power-law Fit} \\
\cline{2-4}\cline{6-9}\\
& \colhead{$\alpha $} & \colhead{F$_{\nu2}0$} & \colhead{Reduced
 $\chi^2$} & & \colhead{$\alpha $} & \colhead{F$_{\nu2}0$} & \colhead{$\beta$} &
\colhead{Reduced $\chi^2$}\\
}

\startdata
J0105$-$0050 & 1.057 $\pm$ 0.060 & 0.113 $\pm$ 0.008 & 1.70 && 0.310 $\pm$ \, 0.023 & \ \ 0.228 $\pm$ 0.004 & 0.194 $\pm$ \, 0.029 & 1.23 \\
J0105$+$0040 & 0.910 $\pm$ 0.025 & 0.090 $\pm$ 0.007 & 1.86 && 0.589 $\pm$ \, 0.037 & \ \ 0.267 $\pm$ 0.012 & 0.350 $\pm$ \, 0.049 & 1.47 \\
J0136$-$0046 & 1.082 $\pm$ 0.132 & 0.107 $\pm$ 0.019 & 1.81 && 1.683 $\pm$ \, 3.412 & $-$0.145 $\pm$ 2.779 & 2.569 $\pm$ 17.297 & 1.84 \\
J0312$-$0113 & 1.155 $\pm$ 0.047 & 0.040 $\pm$ 0.005 & 0.74 && 0.466 $\pm$ 22.985 & $-$0.680 $\pm$ 7.208 & 6.960 $\pm$ 59.681 & 0.72 \\
J0346$-$0042 & 0.647 $\pm$ 0.039 & 0.193 $\pm$ 0.015 & 1.26 && 0.657 $\pm$ \, 0.660 & $-$0.016 $\pm$ 1.050 & 1.810 $\pm$ \, 4.048 & 1.29 \\
J2045$-$0051 & 0.865 $\pm$ 0.022 & 0.090 $\pm$ 0.004 & 2.71 && 0.903 $\pm$ \, 0.307 & \ \ 0.083 $\pm$ 0.055 & 1.048 $\pm$ \, 0.385 & 2.76 \\
J2111$+$0024 & 0.992 $\pm$ 0.074 & 0.028 $\pm$ 0.016 & 0.57 && 0.427 $\pm$ \, 0.121 & \ \ 0.199 $\pm$ 0.030 & 0.221 $\pm$ \, 0.141 & 0.53 \\
J2119$+$0032 & 1.112 $\pm$ 0.069 & 0.041 $\pm$ 0.006 & 0.83 && 4.180 $\pm$ 16.912 & $-$0.127 $\pm$ 1.163 & 2.754 $\pm$ 12.147 & 0.80 \\
J2123$-$0050 & 1.453 $\pm$ 0.116 & 0.507 $\pm$ 0.045 & 0.69 && 1.565 $\pm$ \, 0.167 & \ \ 0.731 $\pm$ 0.559 & 0.598 $\pm$ \, 1.005 & 0.70 \\
J2134$+$0048 & 1.131 $\pm$ 0.039 & 0.023 $\pm$ 0.003 & 0.97 && 0.905 $\pm$ \, 0.435 & \ \ 0.032 $\pm$ 0.021 & 0.868 $\pm$ \, 0.279 & 0.99 \\
\enddata
\end{deluxetable}

\clearpage

\begin{deluxetable}{lcccccc}
\tabletypesize{\footnotesize}
\tablecaption{Estimation of Host Galaxy Flux\label{t_host}}
\tablewidth{0pt}
\tablehead{
\colhead{Object} & \colhead{$F_{\nu1}$(E)} & \colhead{$F_{\nu2}$(E)} & \colhead{$F_{\nu1}$(Sbc)} & \colhead{$F_{\nu2}$(Sbc)} 
& \colhead{$F_{\nu1}$(Im)} & \colhead{$F_{\nu2}$(Im)} \\
}

\startdata
J0105$-$0050 & 6.7E$-$07 & 4.1E$-$05 & 6.4E$-$04 & 4.4E$-$03 & 3.5E$-$03 & 2.1E$-$02 \\
J0105$+$0040 & 1.8E$-$06 & 7.0E$-$04 & 1.4E$-$03 & 1.3E$-$02 & 9.0E$-$03 & 7.8E$-$02 \\
J0136$-$0046 & 4.6E$-$07 & 9.3E$-$05 & 3.5E$-$04 & 2.4E$-$03 & 2.3E$-$03 & 1.2E$-$02 \\
J0312$-$0113 & 2.3E$-$07 & 1.0E$-$06 & 2.2E$-$04 & 6.0E$-$04 & 1.2E$-$03 & 3.0E$-$03 \\
J0346$-$0042 & 7.5E$-$07 & 6.5E$-$05 & 7.2E$-$04 & 5.0E$-$03 & 3.9E$-$03 & 2.4E$-$02 \\
J2045$-$0051 & 5.4E$-$07 & 3.3E$-$05 & 5.1E$-$04 & 3.4E$-$03 & 2.8E$-$03 & 1.6E$-$02 \\
J2111$+$0024 & 7.9E$-$07 & 3.7E$-$06 & 7.6E$-$04 & 2.2E$-$03 & 4.2E$-$03 & 1.1E$-$02 \\
J2119$+$0032 & 1.5E$-$07 & 6.8E$-$07 & 1.4E$-$04 & 4.0E$-$04 & 7.6E$-$04 & 2.0E$-$03 \\
J2123$-$0050 & 2.0E$-$06 & 9.2E$-$06 & 1.9E$-$03 & 5.4E$-$03 & 1.1E$-$02 & 2.7E$-$02 \\
J2134$+$0048 & 9.9E$-$07 & 4.7E$-$06 & 9.5E$-$04 & 2.7E$-$03 & 5.2E$-$03 & 1.4E$-$02 \\
\enddata
\tablecomments{Flux in units of mJy.} 
\end{deluxetable}

\clearpage

\begin{deluxetable}{lcccccc}
\tabletypesize{\footnotesize}
\tablecaption{Small Blue Bump Flux\label{t_sbb}}
\tablewidth{0pt}
\tablehead{
\colhead{Object} & \colhead{$F_{\nu2}$(SBB)} & \colhead{$\sigma_{\nu2}$(SBB)\tablenotemark{a}} & \multicolumn{2}{c}{reduced $\chi^2$} \\ 
\cline{4-5}\\
 & & & $\sigma_{\nu2}$(SBB) included\tablenotemark{b} & not included\tablenotemark{c} \\
}

\startdata
J0105$-$0050 & 4.78E$-$02 $\pm$ 3.15E$-$03 & 3.19E$-$03 & 1.41 & 1.70 \\
J0105$+$0040 & 6.46E$-$02 $\pm$ 3.82E$-$03 & 1.04E$-$02 & 0.98 & 1.86 \\
J0136$-$0046 & 1.24E$-$02 $\pm$ 2.87E$-$03 & 7.07E$-$04 & 1.80 & 1.81 \\
J0312$-$0113 & 3.15E$-$03 $\pm$ 1.17E$-$03 & 3.10E$-$04 & 0.73 & 0.74 \\
J0346$-$0042 & 3.58E$-$02 $\pm$ 7.85E$-$03 & 2.50E$-$03 & 1.17 & 1.26 \\
J2045$-$0051 & 3.88E$-$02 $\pm$ 3.73E$-$03 & 4.50E$-$03 & 1.69 & 2.71 \\
J2111$+$0024 & 1.33E$-$02 $\pm$ 1.88E$-$03 & 6.68E$-$04 & 0.56 & 0.57 \\
J2119$+$0032 & 6.50E$-$05 $\pm$ 1.45E$-$03 & 3.56E$-$06 & 0.83 & 0.83 \\
J2123$-$0050 & 5.55E$-$02 $\pm$ 7.60E$-$03 & 2.26E$-$03 & 0.69 & 0.69 \\
J2134$+$0048 & 5.05E$-$03 $\pm$ 7.64E$-$04 & 4.88E$-$04 & 0.96 & 0.97 \\
\enddata
\tablecomments{Flux in units of mJy.} 
\tablenotetext{a}{The estimated 1$\sigma $ scatter of the flux variation of the SBB and \ion{Mg}{2} emission-line component.}
\tablenotetext{b}{Reduced $\chi^2$ of the straight-line fit to flux-to-flux plot, $\sigma$(SBB) included in the error.}
\tablenotetext{c}{Reduced $\chi^2$ of the straight-line fit to flux-to-flux plot
listed in Table \ref{t_fffit}.}
\end{deluxetable}

\clearpage

\begin{deluxetable}{lcccc}
\tabletypesize{\footnotesize}
\tablecaption{Vanden Berk Model Fits to Flux-to-flux Plots \label{t_vanden}}
\tablewidth{0pt}
\tablehead{
\colhead{Object} & \colhead{$\alpha _2$} & \colhead{$\beta _2$} & \multicolumn{2}{c}{Reduced $\chi^2$} \\
\cline{4-5}\\
& & & \colhead{\citet{2004ApJ...601..692V}} & \colhead{Straight-line}\tablenotemark{a} \\
}

\startdata
J0105$-$0050 & 1.164 $\pm$ 0.007 & 1.505 & 1.89 & 1.70 \\
J0105$+$0040 & 1.040 $\pm$ 0.006 & 1.419 & 2.11 & 1.86 \\
J0136$-$0046 & 1.056 $\pm$ 0.008 & 1.375 & 1.82 & 1.81 \\
J0312$-$0113 & 1.539 $\pm$ 0.008 & 1.315 & 0.72 & 0.74 \\
J0346$-$0042 & 0.544 $\pm$ 0.004 & 1.499 & 1.43 & 1.26 \\
J2045$-$0051 & 1.107 $\pm$ 0.005 & 1.506 & 2.82 & 2.71 \\
J2111$+$0024 & 1.369 $\pm$ 0.006 & 1.315 & 0.62 & 0.57 \\
J2119$+$0032 & 1.460 $\pm$ 0.007 & 1.315 & 0.83 & 0.83 \\
J2123$-$0050 & 0.745 $\pm$ 0.003 & 1.314 & 1.07 & 0.69 \\
J2134$+$0048 & 1.805 $\pm$ 0.008 & 1.316 & 0.98 & 0.97 \\
\enddata
\tablenotetext{a}{Reduced $\chi^2$ of the straight-line fit to flux-to-flux plot
listed in Table \ref{t_fffit}.}
\end{deluxetable}

\clearpage
\begin{deluxetable}{lcccc}
\tabletypesize{\footnotesize}
\tablecaption{Standard Accretion Disk Model Fit to Flux-to-flux Plots \label{t_simstat}}
\tablewidth{0pt}
\tablehead{
\colhead{Object} & \colhead{$\log(M_{BH}/M_{\odot})$\tablenotemark{a}} & \multicolumn{2}{c}{Reduced $\chi^2$} & \colhead{Bolometric Correction\tablenotemark{c}} \\
\cline{3-4}\\
 & & \colhead{AD model} & \colhead{Straight-line\tablenotemark{b}} & \\
}

\startdata
J0105$-$0050 & 9.38$^{+0.01}_{-0.01}$ & 1.86 & 1.70 & 6.04 \\
J0105$+$0040 & 9.18$^{+0.02}_{-0.01}$ & 2.03 & 1.86 & 7.00 \\
J0136$-$0046 & 9.70$^{+0.01}_{-0.01}$ & 1.92 & 1.81 & 5.50 \\
J0312$-$0113 & 9.26$^{+0.01}_{-0.01}$ & 0.72 & 0.74 & 7.30 \\
J0346$-$0042 & 9.90$^{+0.01}_{-0.01}$ & 1.25 & 1.26 & 4.80 \\
J2045$-$0051 & 9.46$^{+0.01}_{-0.01}$ & 2.71 & 2.71 & 5.73 \\
J2111$+$0024 & 9.24$^{+0.02}_{-0.02}$ & 0.55 & 0.57 & 8.63 \\
J2119$+$0032 & 9.43$^{+0.01}_{-0.01}$ & 0.93 & 0.83 & 5.95 \\
J2123$-$0050 & 9.84$^{+0.01}_{-0.01}$ & 1.11 & 0.69 & 5.84 \\
J2134$+$0048 & 9.02$^{+0.02}_{-0.02}$ & 1.06 & 0.97 & 8.88 \\
\enddata
\tablenotetext{a}{The black hole mass of the best-fit standard accretion disk model. The errors are estimated from the statistical uncertainty of $\chi^2$ fitting of the model.}
\tablenotetext{b}{Reduced $\chi^2$ of the straight-line fit to flux-to-flux plot
listed in Table \ref{t_fffit}.}
\tablenotetext{c}{The ratio of the bolometric luminosity
of the best-fit standard accretion disk model
to the specific luminosity in UV,
$L_{\rm bol}/\lambda L_{\lambda }(3000{\rm \AA})$.}
\end{deluxetable}

\end{document}